\documentclass{iaus}
\usepackage{graphicx}

\title[Abundances and ADFs in PNe with WC central stars] 
{Abundances and ADFs in PNe with [WC] central stars}

\author[J. Garc\'{\i}a-Rojas, M. Pe\~na, C. Morisset \& M. T. Ruiz]   
{J. Garc\'{\i}a-Rojas$^1$
\thanks{Based on data collected at Las Campanas Observatory, Chile.},
 M. Pe\~na$^2$,
 C. Morisset$^{1,2}$
 \and M.~T. Ruiz$^3$}

\affiliation{$^1$IAC, E-38200, La Laguna, Tenerife, Spain. email: {\tt jogarcia@iac.es} \\[\affilskip]
$^2$IA-UNAM. Apdo Postal 70264, Mex. D. F., 04510 Mexico. email: {\tt miriam@astroscu.unam.mx, chris.morisset@gmail.com} \\[\affilskip]
$^3$DAS. U. Chile. Casilla 36D, Santiago, Chile. email: {\tt mtruiz@das.uchile.cl} }

\pubyear{2011}
\volume{283}  
\pagerange{}
\jname{Planetary Nebulae: An Eye to the Future}
\editors{A. Manchado, \& L. Stanghellini, eds.}
\begin{document}

\maketitle

\begin{abstract}
We present preliminary results obtained from the analysis of very deep echelle spectra 
of a dozen planetary nebulae with [WC] or weak emission lines (wels) central stars. 
The computed abundance discrepancy factors (ADFs) are moderate, with values lower than 4. 
In principle, no evidence of the H-poor metal enriched inclusions proposed by \cite[Liu et al. 
(2000)]{liuetal00} have been found. However, a detailed analysis of the data is in progress. 
\end{abstract}

\firstsection 
\section{Introduction and Observations}

PNe around [WR] central stars constitute a particular photoionized nebula class, 
representing about 10-15\% of the PNe with known progenitor. In principle, these PNe 
seem suitable for analyzing the abundance discrepancy found when computing abundances from 
faint recombination lines (RLs) or from collisional excited lines (CELs).
\cite[Liu et al. (2000)]{liuetal00} proposed the presence 
of H-deficient knots embeded in the hot plasma as responsible of large ADFs (i.e. RLs/CELs abundance ratios).
These knots could be plausible around H-poor WR stars. 
High spectral resolution spectra were obtained with MIKE echelle spectrograph in the 
6.5m Clay Magellan Telescope at Las Campanas Observatory (Chile) The spectral resolution 
varied from $\sim$10.8 km s$^{-1}$ in the blue to $\sim$12.8 km s$^{-1}$ in the red. Data were 
reduced and flux calibrated. 

\section{Plasma Diagnostics and Chemical abundances}

We have obtained physical conditions, $T_e$ and $n_e$, from several 
diagnostic line ratios by using state-of-the-art atomic data and a preliminary version of 
the package PyNEB of IRAF (see Luridiana et al. 2011). We have constructed 
$T_e$ vs. $n_e$ diagnostic plots for each object. 
In general, we have found that $n_e$([Ar IV]) $>$ $n_e$([Cl III]) $>$ $n_e$([S II]) $>$ 
$n_e$([OII]). Low-ionization electron temperature, $T_e$([N II]), was corrected from the effect of 
recombination contribution to the auroral [N II] $\lambda$5755 line. In Figure 1a. we show 
the behavior of $T_e$([O III])/$T_e$([N II]) ratio with the O$^{++}$/O$^+$. For 
comparison we include data from \cite[Pe\~na et al. (2001)]{penaetal01} $-$WR PNe, red$-$ and 
from \cite[Gorny et al. (2009)]{gornyetal09} $-$Non WR PNe, blue$-$. There is a tendency to lower 
temperature ratio with the increasing O$^{++}$/O$^+$. This tendency could not be 
explained by photoionization models and should be investigated further.

Ionic abundances of several heavy metal ions were computed from CELs. We used $T_e$([N II]), 
$T_e$([O III]) and $T_e$([Ar IV]), when available for the low, intermediate and high ionization 
zones, respectively. Total abundances were obtained using the set of ICFs 
proposed by \cite[Kingsburgh \& Barlow (1994)]{kb94}. Since most of the ICFs depend on the ionization 
degree (O$^+$/O) which in turn depends strongly on the adopted $n_e$, we have explored 
the behavior of Ne/H ratio $vs.$ O/H ratio in Figure 1b. by using an averaged $n_e$ (open 
triangles) or by assuming $n_e$([Cl III]) as characteristic of the whole nebula (filled triangles). 
The dispersion in the relationship between Ne/H (which is an $\alpha$ element) and 
O/H is clearly lower when adopting $n_e$([Cl III]), and this effect is especially 
important in the high density objects, where [O II] and [S II] $n_e$ diagnostics are saturated. 
Two objects, whatever density is used, have very low neon abundance. 
A more detailed abundance study is in progress.

\begin{figure}[!htb]
\begin{center}
 \includegraphics[width=4.1in]{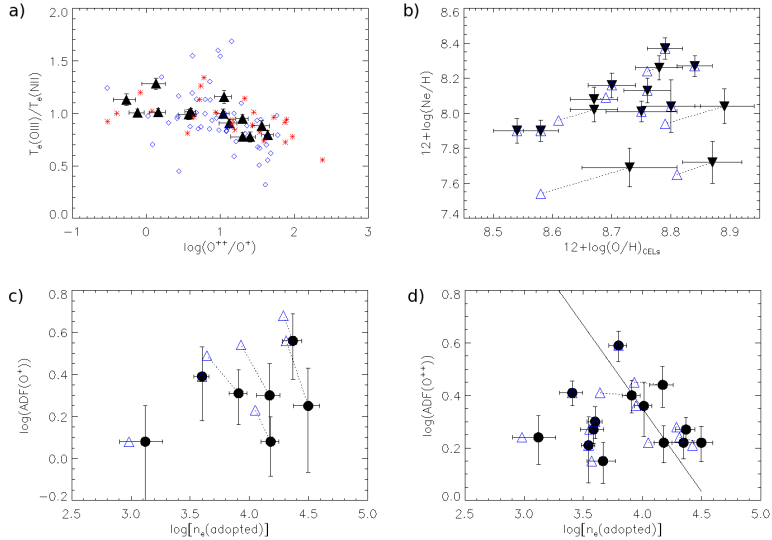} 
 \caption{a) Temperature ratio $vs.$ ionization degree. b) Ne/H $vs.$ O/H. c) ADF(O$^+$) $vs$ $n_e$ d) ADF(O$^{++}$) $vs.$ $n_e$. 
 The relationship found by \cite{rg05} is shown.}
   \label{fig1}
\end{center}
\end{figure}

\section{Abundance Discrepancies}

The ADF was computed for O$^+$/H$^+$ and 
O$^{++}$/H$^+$ ratios. In Figure 1c. and 1d. we show the behavior of both ADFs 
with the adopted $n_e$, respectively. The influence of the adopted density (averaged, 
open symbols) or $n_e$([Cl III]) (filled symbols) is also showed. It is clear the influence 
of the adopted density on the computed ADF(O$^+$). This effect is very small in the derived 
ADFs(O$^{++}$). From that Figure we can also see that ADFs are 
always moderate for both ions (O$^+$ and O$^{++}$) in these objects. 
No direct evidence for the presence of cold and high density clumps has been found. 


\end{document}